\documentclass[%
preprint,
 amsmath,amssymb,
 aps,
]{revtex4-2}

\usepackage{adjustbox}
\usepackage{physics}
\usepackage{amsmath}
\usepackage{array}
\usepackage{setspace}
\newcolumntype{C}[1]{>{\centering\arraybackslash}m{#1}}
\usepackage{xcolor}
\usepackage{graphicx}% Include figure files
\usepackage{dcolumn}% Align table columns on decimal point
\usepackage{bm}% bold math
\usepackage{caption}
\usepackage{subcaption}
\usepackage{rotating}

\begin{document}

\preprint{APS/123-QED}

% \title{Stochastic \rm{GW} with an optimally tuned range-separated hybrid DFT starting point}

\title{GW with hybrid functionals for large molecular systems}

\author{Tucker Allen}
\email{tuckerallen27@ucla.edu}
\affiliation{Department of Chemistry and Biochemistry, University of California, Los Angeles,
Los Angeles, CA, 90095, USA}
\author{Minh Nguyen}
\affiliation{Department of Chemistry and Biochemistry, University of California, Los Angeles,
Los Angeles, CA, 90095, USA}
\author{Daniel Neuhauser}
\affiliation{Department of Chemistry and Biochemistry, University of California, Los Angeles,
Los Angeles, CA, 90095, USA}

\begin{abstract}
A low-cost approach for stochastically sampling static exchange during TDHF-type propagation is presented. This enables the use of an excellent hybrid DFT starting point for stochastic \rm{GW} quasiparticle energy calculations. 
Generalized Kohn-Sham molecular orbitals and energies, rather than those of a local-DFT calculation, are used for building the Green's function and effective Coulomb interaction. The use of an optimally tuned hybrid diminishes the starting point dependency in one-shot stochastic \rm{GW}, effectively avoiding the need for self-consistent \rm{GW} iterations. 
%mention improved fundametnal bandgaps but not IPs (homos) 
\end{abstract}
\maketitle

\section{Introduction}
Kohn-Sham density functional theory (DFT) with local and semi-local functionals has been successful in calculating ground state energies and configurations, but is insufficient for processes that involve excited states such as (inverse) photoemission spectroscopy, as it gives too-small bandgaps and incorrect quasiparticle (QP) energies.~\citep{perdew1985density} An established alternative is the \rm{GW} approximation to many-body perturbation theory (MBPT) for \emph{ab-initio} description of QPs.~\citep{Gunnarsson_1998,Onida_RMP_2002,Reining_2018}
%In essence, the method replaces the approximate exchange-correlation (XC) potential of density-functional theory (DFT) with the Fock exchange of Hartree-Fock (HF) and a screened Coulomb, rather than bare, interaction. 

The \rm{GW} method approximates the single-particle self-energy, which contains all many-body effects, as $\Sigma \approx iGW$, where $G$ is the Green's function that gives the probability amplitude of a QP to propagate between two space-time points, and $W$ is the screened Coulomb interaction.
Common implementations of \rm{GW} perturbatively correct the mean-field orbital energies, providing significant improvements; however, the method's perturbative nature yields observables that strongly depend on the initial set of mean-field orbitals and energies. ~\citep{isseroff2012importance,Bruneval_2013} Therefore, Hedin's \rm{GW} equations should ideally be solved iteratively.~\citep{Hedin_1965} 

Neglecting the vertex function, the \rm{GW} method begins with the noninteracting Green's function, $G_0$, built with mean-field orbitals, and then one computes the screened interaction $W$ with these same orbitals to obtain the self-energy $\Sigma$. In iterative \rm{GW}, the dressed Green's function $G=G_0+G_0\Sigma G$ is then self-consistently updated till convergence. In contrast, one-shot $\rm{G_0W_0}$ avoids updating the Green's function and uses the noninteracting form. The least costly form of partial self-consistency, eigenvalue self-consistent $\rm{GW_0}$ (ev-$\rm{GW_0}$), follows this framework of freezing $W$ and only updating the eigenvalues entering $G$.~\citep{Stan_2009, Vlek2018scissors} 

%The screened interaction $W$ is only computed once, and 

Many groups have studied the performance of \rm{GW} ranging from partial to full self-consistency.~\citep{Blase_GW_wfock2011,Chelikowsky_2012,Bruneval2014,Chen_scGW2014} Eigenvalue-only schemes improve {$\rm{G_0W_0}$} results, and QP (qp-\rm{GW}) self-consistency can help further remove the starting point dependency in the QP energies.~\citep{Kotani_qpsc_2004,Setten2016qpsc} However, qp-\rm{GW} can be very expensive, though a low-scaling qp-\rm{GW} well suited for molecules has been developed.~\citep{forster_lowscalingGW2021,adf_forster2023} Only fully self-consistent \rm{GW} (sc-\rm{GW}) formally completely removes the starting point dependency and other ailments of a first-order perturbation theory.~\citep{Caruso_locGW2013, Kresse_fullscGW2018} Recent work has found that a faithful description of molecular ionization potentials (IPs) can be achieved with sc-\rm{GW}.~\citep{zgid_scGW_IPs} Further, sc-\rm{GW} gives access to thermodynamic quantities, including total ground- and excited-state energies as well as the electron density. However, iteratively solving the Dyson equation is prohibitively expensive for large systems.~\cite{HolmBarth_1998,Valentin2014FirstPA} 

For molecular systems with hundreds to thousands of electrons, a more realistic strategy is to improve the mean-field starting point within $\rm{G_0W_0}$. A hybrid functional with exact exchange starting point improves the accuracy of a one-shot $\rm{G_0W_0}$, and in practice, the results of hybrid-DFT+$\rm{G_0W_0}$ calculations become almost equivalent to ev-$\rm{GW_0}$.~\citep{Caruso_hybridGW2013,Bruneval_2013} Specifically, amongst local, semi-local, global and RSH-DFT starting points, use of an optimally-tuned (OT)-RSH-DFT starting point has been shown to give the lowest mean absolute errors (MAE) relative to experiment.~\citep{Neaton_OTRSH_GW2022,McKeon2022} 

We have developed an efficient hybrid DFT method utilizing sparse-stochastic compression of the exchange kernel.~\citep{neargap_2023} (Note that sparse-stochastic compression was first developed for stochastic \rm{GW} (s\rm{GW}) to reduce memory costs associated with time-ordering the retarded effective interaction $W^R$.~\citep{neuhauser2014breaking,vlvcek2017stochastic,vlvcek2018swiftg}) For hybrid DFT, the exchange kernel is fragmented in momentum space with the vast majority of wave-vectors $k$ represented through a small basis of sparse-stochastic vectors, except for a small fixed number of long-wavelength (low-$k$) components that are evaluated deterministically. These hybrid eigenstates, energies, and Hamiltonian serve as the DFT starting point of the present method.  

%low-$k$ wave-vectors treated deterministically and the large remainder of high-$k$ wave-vectors are represented with a small basis of sparse-stochastic vectors. 

Fully stochastic OT-RSH-DFT was previously used as an alternative to \rm{GW} for frontier QP energies of large silicon nanocrystals.~\citep{neuhauser_stochastic_OTRSH} We also developed a tuning procedure for OT hybrids to reproduce the \rm{GW} fundamental bandgaps of periodic solids.~\citep{li2021tuning} The present hybrid DFT starting point has very tiny stochastic error and differs from previous stochastic methods by avoiding stochastic sampling of the orbitals or density matrix.

%(For the remainder of the paper, the terms GKS, Hybrid, and OT-RSH are used interchangeably).

%Specifically, the underlying GKS Hamiltonian is used for constructing and propagating the noninteracting Green's function, and 

%The stochastic paradigm for $G_0W_0$ (labeled s\rm{GW}) enables a linear scaling method for accurate QPs energies, including ionization potentials (IP) and electron affinities (EA).~\cite{neuhauser2014breaking,vlvcek2017stochastic,vlvcek2018swiftg} The approach combines several novel stochastic techniques, including the stochastic resolution of the identity, sparse-stochastic compression, stochastic decoupling of space-time coordinates, and stochastic time-dependent Hartree (sTDH) propagation. Our form of ev-$GW_0$, $\bar{\Delta} GW_0$ ~\citep{Vlek2018scissors}, trivially modifies the time-dependent self-energy of an s\rm{GW} calculation and amounts to a rigid-scissors shift of the respective occupied and unoccupied subspaces at negligible additional cost. This simplified version of ev-$GW_0$ agrees well with experiment, and the hybrid starting point s\rm{GW} method presented in this article produces fundamental bandgaps unchanged after passing through a $\bar{\Delta} GW_0$ SCF cycle.  

In this article, s\rm{GW} is extended to hybrid functional starting points. The s\rm{GW} algorithm is briefly reviewed and we emphasize specific parts of the method that have been modified for employing hybrid rather than local exchange-correlation (XC) functionals. Additionally, we implement a ``cleaning'' procedure during time-propagation to avoid numerical instabilities, as we have recently done in generating a stochastic $W$ for Bethe-Salpeter equation (BSE) spectra.\citep{Bradbury2022,bradbury_bse_2023} As demonstrated in Ref.~\citep{vertex2019_vojtech}, stochastic fluctuations worsen for hybrid DFT (compared to local/semi-local XC) starting-point $\rm{G_0W_0}$ calculations because of the added sampling of the static exchange. The cleaning procedure significantly reduces the stochastic error in QP energies and enables routine hybrid starting-point $\rm{G_0W_0}$ for molecular systems with thousands of valence electrons. The new approach is tested on a set of molecules with an OT-RSH functional. For technical details of the s\rm{GW} method, we refer to Refs. ~\citep{neuhauser2014breaking,vlvcek2017stochastic,vlvcek2018swiftg}. 

\section{\label{sec:Methodology}Theory}

The stochastic paradigm for $\rm{G_0W_0}$ uses the space-time representation of the single-particle self-energy 
\begin{equation}
    \Sigma(r,r',t) = iG_0(r,r',t)W_0(r,r',t^+)
\end{equation}
which takes a simple direct product form of the noninteracting single-particle Green's function $G_0$ and screened Coulomb  interaction $W_0$. In Section A, we provide a stochastic form of $G_0$ and formulate how to obtain QP energies with a hybrid DFT basis. The sampling of static exchange during propagation of the Green's function is also discussed. 

It is convenient to split the self-energy as $\Sigma=\Sigma_\text{X}+\Sigma_\text{P}$, with an instantaneous Fock exchange part and time-dependent polarization part, respectively. $\Sigma_\text{X}$ is evaluated deterministically and $\Sigma_\text{P}$ is evaluated by a stochastic linear-response time-dependent Hartree propagation (sTDH) ~\citep{neuhauser2014breaking}, equivalent to the standard random phase approximation (RPA). In Section B, the sTDH approach is presented and we introduce a new projection routine that reduces statistical fluctuations when evaluating the action of $W_0$ on a source term. 

\subsection{Stochastic {\rm{GW}} in space-time domain}
\subsubsection{Single-particle Green's function}

In s\rm{GW}, the single-particle Green's function is converted to a random averaged correlation function. We first define the operator form of the zero-order Generalized Kohn-Sham (GKS) Green's function: 
\begin{equation}
\label{G_operator}
    iG_0(t)=e^{-iH_0t}[(\emph{I}-\emph{P})\theta(t)-\emph{P}\theta(-t)], 
\end{equation}
where \emph{I} is the identity matrix and $P=\sum_{n\le{N_\text{occ}}} |\phi_n\rangle \langle \phi_n|$ projects onto the occupied subspace of the ground state GKS Hamiltonian
\begin{equation}
    H_0 = -\frac{1}{2}\nabla^2 + v_{\text{e-N}} + v_\text{H}[n_0]+v_\text{XC}^\gamma[n_0]+X^\gamma[\rho_0].
\end{equation}
Assuming closed-shell systems, the static density matrix is
\begin{equation}
    \rho_0(r,r')=2\sum_{n\le N_{\text{occ}}} \phi_n(r)\phi_n^*(r'),
\end{equation}
and the diagonal elements yield the density $n_0(r)=\rho_0(r,r'=r)$. $H_0$ includes the kinetic energy, overall electron-nuclear potential, Hartree potential, exchange-correlation potential, and long-range exact exchange. The range-separation parameter $\gamma$ divides the exchange interaction into short- and long-range components. For the short-range part a local (or semi-local) exchange energy is used, while for the long-range part  a parameterized exact exchange operator is used.~\citep{savin1997,Becke1993exact} The range-separation parameter is optimally tuned to each system of interest, and the tuning procedure enforces the ionization potential theorem of DFT.~\citep{baer_tuned_2010}

Stochastic projection onto the occupied and unoccupied subspaces (as in Refs.~\citep{vlvcek2018swiftg,Vlek2018scissors}) begins with $N_\zeta$ different discrete random vectors, each with entries $\zeta(r)=\pm 1/\sqrt{dV}$ at each grid-point, where $dV$ is the volume element. These vectors form a stochastic resolution of the identity $I\simeq\frac{1}{N_\zeta} \sum_{\zeta}|\zeta\rangle\langle\zeta|$. Inserting this identity to Eq. (\ref{G_operator}) produces 
\begin{equation}
    G_0(r,r',t) \simeq \frac{1}{N_\zeta} \sum_{\zeta} \tilde{\zeta}(r,t) \zeta^*(r')
\end{equation}
where $\tilde{\zeta}(r,t)$ is divided into positive (negative) times corresponding to the propagation of electrons (holes): 
\[
 \tilde{\zeta}(r,t)  \equiv
\begin{cases} 
      \langle r|e^{-iH_0t}P|\zeta\rangle & t < 0 \\
      \langle r|e^{-iH_0t}(I-P)|\zeta\rangle & t > 0 
   \end{cases}
\]
In contrast to deterministic methods where exact eigenstates are evolved in time, a stochastic evaluation of the Green's function requires propagation of randomly projected states; this makes the extension to hybrid functional starting points non-trivial. We discuss how to efficiently apply $X^\gamma[\rho_0]$ at each time-step in Sec. \ref{actionX}. 

\subsubsection{QP energies  with hybrid functionals}

The input GKS MOs and energies fulfill $H_0 \phi_j=\varepsilon_j \phi_j$. In this section, the GKS orbital energy is expressed as a sum of three parts: 
\begin{equation}
    \varepsilon \equiv \langle \phi| H_\text{non-xc}+v_\text{XC}^\gamma+ X^\gamma| \phi\rangle,
\end{equation}
where 
\begin{equation}
    H_\text{non-xc} = -\frac{1}{2}\nabla^2 + v_{\text{e-N}} + v_\text{H}[n_0].
\end{equation}
In a one-shot calculation, the diagonal \rm{GW} QP energy is then perturbatively evaluated as: 
\begin{equation}
\label{gw_pert}
\varepsilon^\text{QP} =\langle \phi| H_\text{non-xc}+\Sigma_\text{X}+\Sigma_\text{P}(\omega=\varepsilon^\text{QP})|\phi\rangle,
\end{equation}
where $\Sigma_\text{X}$ is the Fock exchange self-energy operator with diagonal element for orbital $\phi$:
\begin{equation}
    \langle \phi|\Sigma_\text{X}|\phi\rangle = - \int \frac{\phi^*(r)\phi(r')}{|r-r'|}\rho_0(r,r')dr'dr, 
\end{equation}
and $\Sigma_\text{P}$ is the polarization self-energy operator.

%With a GKS-DFT starting point, the QP energy is: 
%\begin{equation}
%    \varepsilon^\text{QP} = \varepsilon + \langle \phi| X-X^\gamma+\Sigma_\text{P}(\omega=\varepsilon^\text{QP})-v^\gamma_{\text{XC}}|\phi\rangle.
%\end{equation}
To avoid re-evaluating the approximate long-range exact exchange $X^\gamma$ already done once in the hybrid DFT stage, we write
\begin{equation}
    \varepsilon^\text{QP} = \varepsilon + \Delta,
\end{equation}
where 
\begin{equation}
    \Delta = -\delta\varepsilon_0 + \langle \phi|\Sigma_\text{X}+\Sigma_\text{P}(\omega=\varepsilon^\text{QP})|\phi\rangle,
\end{equation}
and $\delta\varepsilon_0\equiv\varepsilon-\langle\phi|H_\text{non-xc}|\phi\rangle$ contains the approximate hybrid long-range exchange and short-range LDA XC potential from the GKS calculation. 

\subsubsection{Stochastic sampling of static exchange}
\label{actionX}

The GKS Hamiltonian (including $X^\gamma[\rho_0]$) must be applied at every time-step when acting with $G_0(t)$ and $W_0(t)$. Below, we detail the stochastic sampling for the Green's function. This is done analogously for the sTDH stage. 
The action of the long-range exchange operator $X^\gamma[\rho_0]$ in the static $H_0$ is approximated very simply as: 
\begin{equation}
    e^{-iX^{\gamma}dt} \approx C_{\text{norm}}(1-iX^{\gamma}dt), 
\end{equation}
where $C_{\text{norm}}$ is a normalization factor to conserve the norm of the propagated states. The action of $e^{-iX^{\gamma}dt}$ must be repeatedly evaluated at every time-step, and to do this we develop an improvement of earlier approaches. A previous approach to implement long-range explicit exchange starting points for s\rm{GW} ~\citep{vertex2019_vojtech} sampled the same set of stochastic orbitals used for propagation, resulting in increased stochastic noise. This issue also appeared in early work on the stochastic BSE approach for optical spectra.~\citep{Rabani_2015}

%In this work, the static long-range exchange is calculated by first introducing a small set of stochastic vectors that sample the deterministic ground state occupied orbitals. This set is then repeatedly sampled at every time-step to approximate the full occupied orbital subspace by a single stochastic vector. 

To overcome these deficiencies, we first introduce an intermediate basis of $N_\beta$ random functions 
\begin{equation}
\label{beta_nu}
    \beta_\nu(r) = \sum_{i\in N_{occ}} a_{i\nu} \phi_i(r),
\end{equation}
with coefficients
\begin{equation}
    a_{i\nu} = \frac{1}{N_\beta^{1/2}} e^{\mathfrak{i}\theta}, \theta \in [0,2\pi]. 
\end{equation} 
The coefficients draw a random phase from the complex unit circle, and give an equal amplitude for each $\phi_i$ orbital in the summation. These functions randomly scramble the information of the occupied subspace, and give an approximate ground state density matrix
\begin{equation}
    p_{0}(r,r')\simeq \sum_{{\nu\in}N_\beta} \beta_\nu(r)\beta_\nu^*(r').
\end{equation}
We then prepare at every time-step a \emph{new} random vector, 
\begin{equation} u(r,t)=\sum_{{\nu\in}N_\beta} \pm\beta_\nu(r).
\end{equation}
This vector is an instantaneous random linear combination of the finite $\beta_\nu$ functions that themselves stochastically sample the occupied space. 

Since the exchange operator is time-independent, the $\beta_\nu$  functions do not need to be updated at every time-step. This differs from previous stochastic TDHF (time-dependent Hartree Fock) approaches where $X^\gamma$ is based on $\rho(t)$ rather than $\rho_0$.~\citep{Rabani_2015} Sampling all the orbitals at every time-step would be expensive so we use this intermediate step of the $\beta_\nu$ functions. The required number of random occupied functions is quite small, with $N_\beta\approx50$ used in this work. Convergence with this parameter is discussed in the results section. 

Now, we are able to evaluate the action of the long-range exchange operator on an arbitrary state $\psi$:
\begin{equation}
\label{eq:actionX}
    \langle r|X^\gamma|\psi\rangle = - u(r,t)\int v_\gamma(|r-r'|)u^*(r',t)\psi(r')dr',
\end{equation}
where $v_\gamma(|r-r'|)=\text{erf}(\gamma|r-r'|)/{|r-r'|}$ is the long-range exchange kernel. This stochastic sampling removes the sum over occupied states that appears when evaluating the action of exact exchange on a general ket.

\subsection{Obtaining the polarization self-energy $\Sigma_\text{P}$ }
\subsubsection{Stochastic TD-Hartree propagation}
\label{sTDH}

%Inserting a stochastic resolution of the identity

The polarization self-energy is evaluated on the real-time axis through the causal (retarded) linear response to an external test charge. Deterministically, this would be done through a time-dependent Hartree (TDH) propagation. However, TDH is prohibitive for large systems as one has to propagate all occupied orbitals. 
The stochastic approach circumvents this problem by perturbing and propagating a small number of orbitals that are each a random linear combination of all occupied states. 
Here we outline the parts of the sTDH propagation that relate to using an underlying GKS Hamiltonian. For a more detailed derivation of sTDH, we refer to Ref. \cite{vlvcek2018swiftg}.

%In the time-domain, the diagonal expectation value of the polarization self-energy is
%\begin{equation}
%    \langle\phi|\Sigma_\text{P}|\phi\rangle \simeq \frac{1}{N_\zeta} \int \phi(r) \zeta(r,t)u(r,t)
%\end{equation}

For each vector $\zeta$, we define a small set of $N_\eta(\approx 10-20$) stochastic orbitals
\begin{equation}
    \eta_{l}(r) = \sum_{i\in N_{occ}} \eta_{il}\phi_i(r),
\end{equation}
where $\eta_{il}=\pm 1$ and $\{\phi_i\}$ are the occupied GKS eigenfunctions. We emphasize that these $\eta_{il}$ coefficients are different than the $a_{i\nu}$ coefficients for sampling the static exchange. The use of numerically independent random bases helps to avoid numerical bias in our results. 

For orbital $\phi_n$, computing the diagonal element of the polarization self-energy, $\langle \phi_n|\Sigma_\text{P}(t)|\phi_n\rangle$, requires that these stochastic occupied functions will be perturbed 
\begin{equation}
    \eta^{\lambda}_l(r,t=0)=e^{-i\lambda v_{\text{pert}}} \eta_l(r)
\end{equation}
with a Hartree-like potential $v_{\text{pert}}(r)\equiv \int |r-r'|^{-1} \zeta(r')\phi_n(r')dr'$, where $\lambda$ is a perturbation strength $\approx 10^{-4}$. These states are then evolved $\eta^{\lambda}_l(r,t+dt)=e^{-iH^{\lambda}(t)dt}\eta^{\lambda}_l(r,t)$ under the RPA (i.e., TDH) Hamiltonian
\begin{equation}
    H^{\lambda}(t) = H_0 + v_\text{H}[n^{\lambda}(t)] - v_\text{H}[n_0], 
\end{equation}
with the split-operator technique. Here $n^{\lambda}(r,t)\equiv \frac{1}{N_\eta}\sum_l |\eta^{\lambda}_l(r,t)|^2$, and note that an additional set of unperturbed $\lambda=0$ stochastic orbitals must also be propagated. From the potential difference one obtains the retarded response: 
\begin{equation}
    u^R(r,t)=\frac{1}{\lambda} (v_H[n^{\lambda}(r,t)]-v_H[n^{\lambda=0}(r,t)]).
\end{equation}
This time-dependent potential accounts for the variation in the Hartree field due to the introduction of a QP at $t=0$. Through some manipulations (see Ref.~\citep{neuhauser2014breaking}), $u^R(r,t)$ yields the time-ordered action of $W_0(t)$ on the stochastic source term $\zeta\phi_n$, eventually yielding the \rm{GW} polarization self-energy.   

\subsubsection{Orthogonality routine}
Numerical instabilities during sTDH propagation may occur due to the contamination of the excited component $\eta_l^\lambda(t)-\eta_l^{\lambda=0}(t)$ by occupied state amplitudes'. This artifact of the stochastic method 
is greatly alleviated by a method we introduced earlier in our BSE work (see Ref. \cite{Bradbury2022}), i.e., periodically ``cleaning'' the stochastic orbitals. More specifically, at every $M^{\text{th}}$ time-step, enforce orthogonality of the stochastic perturbed propagated states to all GKS occupied states by projecting onto the virtual subspace: 
\begin{equation}
      \eta^{\lambda}_l(t) \to \eta^{\lambda=0}_l(t) + (I-P)(\eta^{\lambda}_l(t) - \eta^{\lambda=0}_l(t)), 
\end{equation}
and then re-normalizing the  
``cleaned'' $\eta_l^{\lambda}(t)$ vectors. For the present simulations, this projection is done every $M=10$ time-steps, where $dt=0.05$, and the overall simulation time is 50 au. While previous implementations of sTDH in s\rm{GW} propagated the $\eta_l^{\lambda}(t)$ and $\eta_l^{\lambda=0}(t)$ vectors separately, this added projection step requires now the simultaneous propagation of the two sets of states. 

%Note also that the stochastic states introduced above are not the same functions used for generating the time-dependent density and Hartree energy.

%Finally, to linearize we write 
%\begin{equation}
%    \varepsilon^\text{QP} \simeq \varepsilon + \Delta_0 + \frac{\partial\Sigma_\text{P}}{\partial\omega}(\varepsilon^\text{QP}-\varepsilon). 
%\end{equation}
%A few trivial algebra %manipulations then gives 
%\begin{equation}
%    \varepsilon^\text{QP}\simeq %\varepsilon+Z\Delta_0,
%\end{equation}
%with QPs weight $Z=(1-%\frac{\partial\Sigma_\text{P}}%{\partial\omega})^{-1}$.

\section{\label{sec:Results}Results}

We test hybrid-s\rm{GW} on various finite molecules, including urea, a series of polycyclic aromatic hydrocarbons, a model Chlorophyll-a (Chla) monomer dye, and a hexamer photosynthetic dye complex found at the reaction center of Photosystem II (RC-PSII). ~\citep{Forster_ADF_Chromo_2022} 

\begin{table*}
\centering
\begin{adjustbox}{max width=\textwidth}
\begin{tabular}{|l|r|r|r|r|r|r|r|r|r|r|}
\hline
\rule{0pt}{3ex}
\textbf{Molecule} &
  \multicolumn{1}{l|}{$\gamma$ ($\text{Bohr}^{-1}$)} & 
  \multicolumn{1}{l|}{\textbf{LDA-DFT}} &
  \multicolumn{1}{l|}{\textbf{LDA+s\rm{GW}}} &
  \multicolumn{1}{l|}{\textbf{LDA+$\bar{\Delta}\rm{GW}_0$}} &
  \multicolumn{1}{l|}{\textbf{hybrid-DFT}} &
  \multicolumn{1}{l|}{\textbf{hybrid+s\rm{GW}}} &
  \multicolumn{1}{l|}{\textbf{hybrid+$\bar{\Delta}\rm{GW}_0$}} & \textbf{Ref.}
  \\ \hline
Urea         & 0.380 & 4.66 & 9.40 (0.07) & 10.39 (0.07) & 10.32 & 10.58(0.03) & 10.61 (0.03) & \\ \hline
Naphthalene  & 0.285 & 3.34 & 7.60 (0.05) & 7.97 (0.05) & 8.63  & 8.47 (0.04)  & 8.46 (0.04) & 8.73 \\ \hline
Tetracene    & 0.220 & 1.63 & 5.07 (0.03) & 5.35 (0.03) & 5.82  & 5.73 (0.04)  & 5.72 (0.04) & 6.14 \\ \hline
Hexacene     & 0.200 & 0.57 & 3.46 (0.05) & 3.66 (0.05) & 4.26  & 4.05 (0.04)  & 4.04 (0.04) & 4.85 \\ \hline
Chla         & 0.160 & 1.40 & 3.73 (0.06) & 3.88 (0.06) & 4.37  & 4.22 (0.04)  & 4.21 (0.04) & 4.41 \\
\hline
RC-PSII      & 0.120 & 1.23 & 3.82 (0.05) & 3.97 (0.05) & 3.82 & 3.98 (0.05) & 4.00 (0.05) & 4.17 \\
\hline
\end{tabular}
\end{adjustbox}
\caption{Fundamental bandgaps (in eV) using LDA- and hybrid-DFT starting points (with associated $\gamma$) with associated statistical error for various molecules. Literature CCSD(T) values for oligoacenes are from Ref.~\citep{Rangel_oligoacenes2016}. The ev-$GW$ fundamental bandgaps for the Chla monomer and the RC-PSII hexamer are from Ref. \cite{Forster_ADF_Chromo_2022}, which uses a modified version of the ADF 2022 software.~\citep{ADF_software_2001}} 
\label{gaps_table}
\end{table*}

\begin{figure}[]
\centering
\includegraphics[width=0.45\textwidth]{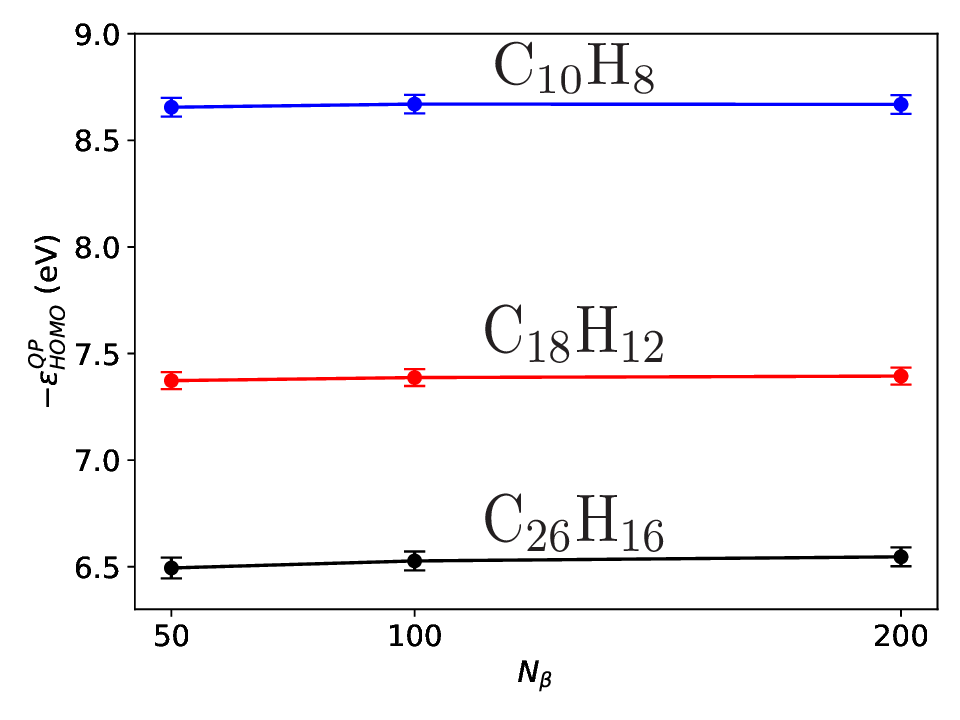}
\label{fig:naph_xc}
\caption{Convergence of the ionization potential (IP) of linear acenes (hybrid-DFT+s\rm{GW}) with respect to $N_\beta$, the number of intermediate stochastic functions for evaluating the static exchange, (Eq. \ref{beta_nu}).}
\end{figure}
%Label $n$ denotes the number of fused benzene rings.

Table \ref{gaps_table} shows the fundamental bandgaps for several approaches: local and hybrid DFT, stochastic \rm{GW} based on an LDA starting point (one-shot and eigenvalue-iterative), and the present hybrid starting-point stochastic \rm{GW} (one-shot and eigenvalue-iterative). The simplified eigenvalue-iterative stochastic \rm{GW}, developed in Ref.~\citep{Vlek2018scissors}, is denoted $\bar{\Delta} GW_0$. The ionization potentials (IP) of the same molecules are shown in Table II. Available reference values are provided for both the fundamental gaps and IPs.

%LDA+s\rm{GW}, LDA+$\bar{\Delta} GW_0$, hybrid-DFT, hybrid+s\rm{GW}, and hybrid+$\bar{\Delta} GW_0$. 

Comparing the LDA and OT-RSH starting points, using the latter raises both the gap and IP by roughly 0.1-0.5 eV. We observe that, except for hexacene, the standalone hybrid-DFT eigenvalues serve as an excellent estimate for the IP and gap, as eigenvalue iterative $\bar{\Delta} GW_0$ barely changes the results. 

For the frontier QP energies of larger acenes, such as hexacene, it is well known~\citep{Rangel_oligoacenes2016} that both one-shot $\rm{G_0W_0}$ and ev-\rm{GW} schemes qualitatively differ from reference CCSD(T) estimates. In Ref.~\citep{vertex2019_vojtech}, only after the addition of a vertex correction was the LUMO QP energy for hexacene sufficiently increased so that the fundamental gap was in excellent agreement with the reference CCSD(T) values of Ref.~\citep{Rangel_oligoacenes2016}. 

The s\rm{GW} calculations introduce correlation via $\Sigma_\text{P}$ which lowers the IPs and gaps of the acenes. However, for urea and the dye systems the s\rm{GW} IPs and gaps are slightly raised relative to the hybrid-DFT.

\begin{table}[h]
\centering
\begin{adjustbox}{max width=\textwidth}
\begin{tabular}{|l|r|r|r|r|r|r|}
\hline
\rule{0pt}{3ex} \textbf{Molecule} & \multicolumn{1}{l|}{\textbf{LDA+s\rm{GW}}} &
  \multicolumn{1}{l|}{\textbf{LDA+$\bar{\Delta}\rm{GW}_0$}} & \multicolumn{1}{l|}{\textbf{hybrid-DFT}} & \multicolumn{1}{l|}{\textbf{hybrid+s\rm{GW}}} & \multicolumn{1}{l|}{\textbf{hybrid+$\bar{\Delta}$\rm{GW}$_0$}} & \textbf{Ref.} \\ \hline
Urea        & 9.35 (0.07) & 10.30 (0.07) & 10.58 & 10.77 (0.05) & 10.79 (0.05) & 10.28 \\ \hline
Naphthalene & 8.21 (0.04) & 8.44 (0.04) & 8.68 & 8.66 (0.03)  & 8.63 (0.03) & 8.25 \\ \hline
Tetracene   & 7.03 (0.02) & 7.22 (0.02) & 7.37 & 7.36 (0.03)  & 7.36 (0.03) & 6.96 \\ \hline
Hexacene    & 6.23 (0.04) & 6.38 (0.04) & 6.59 & 6.50 (0.05)  & 6.49 (0.05) & 6.32 \\ \hline
Chla        & 6.42 (0.06) & 6.58 (0.06) & 6.68  & 6.69 (0.05)   & 6.70 (0.05) & 6.1    \\ \hline
RC-PSII    & 5.76 (0.05) & 5.91 (0.05) & 6.13 & 6.26 (0.05) & 6.27 (0.05) & \\
\hline
\end{tabular}
\end{adjustbox}
\caption{Ionization potentials (in eV) of finite molecules, taken as $I=-\varepsilon^{\text{QP}}_{\text{HOMO}}$, with statistical error of HOMO QP energy. Experimental values for urea and Chla are from Refs. \cite{johnson2006nist,chla_IP_2011}, respectively. High-quality CCSD(T) values for acenes are from Ref. \cite{Rangel_oligoacenes2016}.}
\label{tab:my-table}
\end{table}

Figure 1 shows convergence of the IP of naphthalene, tetracene, and hexacene with respect to the number of $N_\beta$ intermediate-exchange stochastic functions. These are used to sample the action of $X^\gamma[\rho_0]$ on all occupied states during orbital propagation. Going beyond $N_\beta=50$ has minimal effect on the HOMO, and the overall error in the HOMO energy barely changes with $N_\beta$. We have verified for the RC-PSII, with $N_{occ}=660$ occupied orbitals, that increasing $N_\beta$ to $100$ does not change the results. 

Here we review the added computational cost in using hybrid functional starting points for stochastic $\rm{G_0W_0}$. Operation-wise, for each time-step, preparing the vector $u(r,t)$ costs $N_\beta N_g$ operations and the evaluation of the static exchange requires $N_\eta N_g\text{log}N_g$ operations, where $N_g$ is the number of grid points. A technical point is that the convolutions in Eq. (\ref{eq:actionX}) use the Martyna-Tuckerman grid-doubling approach, so the computational effort is higher by an order of magnitude.~\citep{MartynaTuckerman1999} 

We finally showcase the power of this stochastic framework by studying the 1,320 valence electron RC-PSII. The reported OT-RSH-DFT starting point stochastic \rm{GW} QP energies are converged within a statistical error of 0.05 eV using only $N_\zeta=1024$ stochastic runs. Both the nuclear coordinates and reference atomic basis-set ev-\rm{GW} energies for this system are from Ref. \citep{Forster_ADF_Chromo_2022}. The reference calculation uses the PBE global hybrid with $40\%$ exact exchange and includes scalar relativistic effects. We obtain fairly good agreement ($<$0.2 eV discrepancy) between the reference ev-\rm{GW} fundamental bandgap and one-shot s\rm{GW} with an optimally-tuned hybrid starting point. For this large complex, our tuned hybrid has a range-separation parameter of $\gamma=0.12$ $\text{Bohr}^{-1}$. This small value indicates \emph{weak} long-range exchange, and it is reasonable that the s\rm{GW} bandgap of RC-PSII obtained with this starting point is lower than with the global hybrid functional starting point (i.e., the global hybrid method opens the gap more than the tuned hybrid).  

\section{Conclusions\label{sec:conclusion}}
We introduced an approach to efficiently sample static exchange during TDHF-type propagation. This general method is applied here to implement hybrid-DFT starting points to stochastic \rm{GW} calculations. Our results use the long-range Baer-Neuhauser-Lifshitz (BNL) functional,~\citep{baer_density_2005} but the method is amenable to a wide range of hybrids; future studies will benchmark s\rm{GW} with different hybrid-DFT starting points. 

The stochastic sampling approach to evaluate long-range exchange during propagation of $G_0$ and $W_0$ is only $\approx 1.5-2$ times more expensive than using an LDA starting point s\rm{GW}. Because the long-range exchange operator functionally depends on the static density matrix, $\rho_0$, a stochastic sampling approach is the optimal choice as one can sample the occupied subspace only once.

%MAYBE PUT A CARTOON FIGURE SHOWING HOMO-LUMO EIGENVALUE LEVELS GOING FROM LDA TO HYBRID TO GWs

The hybrid s\rm{GW} gaps presented are quite close (usually within less than $0.5$ eV) to the hybrid-DFT gaps, so eigenvalue self-consistency makes only a further minor difference in final bandgaps. One explanation is that the starting GKS energies correspond to a Hamiltonian with the same $-1/r$ asymptotic behavior in the exchange potential as the true \rm{GW} Hamiltonian. This makes GKS energies a much better starting point for the \rm{GW} perturbation expansion (Eq. \ref{gw_pert}) than LDA energies that possess exponential decay in their exchange potential. 

The hybrid-DFT frontier eigenvalues effectively mimic the \rm{GW} HOMO and LUMO QP energies. This holds promise for studying neutral excitations in the stochastic \rm{GW}-BSE (see Refs. ~\citep{Bradbury2022,bradbury_bse_2023}) where the hybrid-DFT wavefunctions will be used for the electron-hole exciton basis, rather than \rm{GW} scissor-corrected local-DFT orbitals or true QP orbitals.

%Chlorophyll a, the hybrid-DFT+s\rm{GW} method improves the fundamental bandgap over LDA+$\bar{\Delta}GW_0$ when compared to a reference all-electron ev-GW method.~\citep{Forster_ADF_Chromo_2022} With a reference gap of $4.41$ eV, LDA+$\bar{\Delta}GW_0$ in Table \ref{gaps_table} has a discrepancy of more than 0.5 eV, while hybrid-DFT+s\rm{GW} agrees within 0.05 eV of the reference value. 

In future work, sparse-stochastic exchange will be applied to vertex corrections of the \rm{GW} self-energy. Vertex corrections have been found to be essential for accurate description of virtual QP energies and plasmons. More fundamentally, inclusion of an approximate non-local vertex can partially correct the self-screening introduced when computing the self-energy in the RPA.~\citep{Romaniello_2009,Romaniello2021} Previous work on low-order stochastic approximations of the vertex function amounts to introducing a scaled non-local exchange term in the polarization part of the self-energy.~\citep{vlcek_vertex2019,gwgamma_vojtech2022,Weng_2023} Our sparse-stochastic exchange technique with orbital-cleaning would efficiently tackle this calculation.

\section*{Acknowledgments}
This work was supported by the U.S. Department of Energy, Office of Science, Office of Advanced Scientific Computing Research, Scientific Discovery through Advanced Computing (SciDAC) program, under Award No. DE-SC0022198. 
This work used the Expanse cluster at San Diego Supercomputer Center through allocation PHY220143 from the Advanced Cyberinfrastructure Coordination Ecosystem: Services \& Support (ACCESS) program \citep{boerner2023access}, which is supported by National Science Foundation grants \#2138259, \#2138286, \#2138307, \#2137603, and \#2138296.

%\section*{Declaration of Competing Interest}
%The authors declare that they have no known competing financial interests or personal relationships financial interests or personal relationships that could have appeared to influence the work reported in this paper.

\section*{Data Availability Statement}
The data that support the findings of this study are available from the corresponding author upon reasonable request.

\bibliographystyle{unsrt}
\bibliography{bib}

\end{document}